\documentclass[desactivate]{aa}  

\usepackage{natbib}
\bibpunct{(}{)}{;}{a}{}{,}

\usepackage{graphicx}
\usepackage{revsymb}	
\usepackage{amsmath}
\usepackage{txfonts}
\usepackage{amssymb}
\usepackage{geometry} 
\usepackage[parfill]{parskip} 
\usepackage{slantsc}
\usepackage{array}
\usepackage{amsfonts}
	
\usepackage{epstopdf}	
\usepackage{epsfig}	

\usepackage{url}

\usepackage[colorlinks=true,linkcolor=black, urlcolor=black, citecolor=black]{hyperref}

\usepackage{color}

\usepackage{xfrac}

\usepackage{sidecap}
\usepackage[font=small, labelfont=bf]{caption}
\usepackage{subcaption}

\usepackage{multirow}
\usepackage{booktabs}
\usepackage{threeparttable}

\begin{document}
   \title{Asteroseismic modelling of \textit{Kepler} Legacy stars including lithium depletion}
\author{G. Buldgen\inst{1} \and J. B\'{e}trisey\inst{2} \and C. Pezzotti\inst{1}  \and S. Borisov\inst{3} \and A. Noels\inst{1}}
\institute{STAR Institute, University of Liège, 19C Allée du 6 Août, 4000 Liège, Belgium \\email: 	\texttt{gbuldgen@uliege.be}
\and Department of Physics and Astronomy, Uppsala University, Box 516, SE-751 20 Uppsala, Sweden
\and D\'epartement d'Astronomie, Universit\'e de Gen\`eve, Chemin Pegasi 51, CH-1290 Versoix, Switzerland}
\date{\today}

\abstract
{The \textit{Kepler} Legacy sample is, to this day, the sample of solar-like oscillators with the most exquisite asteroseismic data. In this work, we carry out a detailed modelling of a subsample of these stars for which the surface lithium abundance has also been observed by the LAMOST survey and a photometric surface rotation has been measured.}
{We aim at studying the impact of additional mixing processes on the asteroseismic modelling of \textit{Kepler} Legacy G and F-type stars. We also investigate whether a single process can be invoked to reproduce the lithium depletion and asteroseismic constraints at the same time}
{We use detailed asteroseismic modelling techniques combining global and local minimization techniques. We start by using standard models and then aim at improving this solution by the addition of extra-mixing at the border of convective regions using either convective penetration or turbulence in radiative layers.}
{We find that lower mass models ($\approx 1 \rm{M}_{\odot}$) have no problem in reproducing the observed lithium depletion using only turbulence in the radiative zone, similarly to solar models. F-type stars, having a shallower convective envelope, are unaffected by additional turbulence at the base of the convective zone, but require significant convective penetration values to actually reproduce the observed lithium depletion. The extent of this penetration is however incompatible with the frequency separation ratios.}
{We conclude that the impact of extra-mixing is moderate for solar-type stars of the \textit{Kepler} Legacy sample and well within the requirements of the PLATO mission. For more massive stars ($\approx 1.5 \rm{M}_{\odot}$), we conclude that the behaviour of the frequency separation ratios must be further investigated, as even models with large convective penetration at the base of their convective envelope are unable to reproduce them.}


\maketitle


\section{Introduction}

Lithium is a key element for stellar evolution, its low fusion temperature (about $2.5\times 10^{6}$K) makes it an extremely sensitive tracer of transport processes happening in stars. Numerous studies have been dedicated to its evolution \citep[e.g.][]{Boesgaard1976,Boesgaard1991,Carlos2019,Carlos2020,Borisov2024a, Borisov2024b}, in an attempt to characterise the underlying physical process at the origin of its depletion \citep[e.g.][]{Baglin1985,Lebreton1987,Montalban1994,Charbonnel1994,Schlattl1999,Montalban2000,Thevenin2017,Baraffe2017,Eggenberger2022,Buldgen2025Li} as well as its link with the presence of exoplanets \citep{Bouvier2008,Castro2009,Deal2015}. It is also a crucial element for primordial nucleosynthesis models \citep[see e.g.][and refs therein]{Fields2011, Clara2020, Deal2021}. A criticial issue is that lithium, unlike beryllium, is significantly burned during the pre-main sequence phase and for a solar-like star, its fusion temperature is actually very close to the temperature of the transition between convective and radiative layers. Therefore the observed lithium depletion can be explained by both turbulence in the radiative layers \citep[e.g. linked with the transport of angular momentum]{Dumont2021,Eggenberger2022} or convective undershooting at the base of the convective envelope. In the solar case, it can be shown \citep[see e.g.][for a recent study]{Buldgen2025Li} that turbulence in the radiative zone is required, as a significant beryllium depletion is observed at the solar age that can only be reproduced by some mixing quite deep in the radiative zone. However, there is no guarantee that the same processes might be acting for other stars, especially as the convective envelope gets thinner with mass. 

From the point of view of angular momentum transport, we know that an efficient process should be acting \citep{Aerts2019}, reducing efficiently the effects of shear induced turbulence. While meridional circulation may lead to efficient transport of chemicals, this effect would be directly correlated to the rotation velocity. Another crucial aspect is the actual efficiency of the magnetic braking for stars with thinner envelopes \citep{Matt2012,Matt2015,Amard2016,Matt2019,Betrisey2023_rot}. 

On the other hand, F-type stars have recently been shown to present glitches in their frequency separation ratios that can be linked with a significantly deeper position of the base of their convective envelope \citep{Lebreton2012,Deal2023,Deal2025}. Such extended convective zones would lead to a much higher depletion of lithium and potentially be traceable in a complementary way to detailed seismic modelling.

In this study, we analyze a sample of 10 solar-like oscillators of G and F spectral type for which high-quality seismic data, surface rotation and surface lithium abundance has been determined. Our work expand on that of \citet{Beck2017} who focused on solar twins, for most of which only global seismic indicators were available. We carry out a first round of modelling using the FICO procedure defined in \citet{Betrisey2023_AMS_surf} and use these models as initial conditions for a supplementary modelling attempting to reproduce the lithium depletion. We start by presenting our sample and the constraints we use in Sect. \ref{sec:sample}. In Sect. \ref{sec:asteroseimic_modelling_using_FICO_procedure}, we detail our initial modelling approach and describe the results of the next modelling step including the observed lithium abundance in Sect. \ref{sec:modelling_including_lithium_depletion}. Finally, in Sect. \ref{sec:origin_mixing_and_how_constrain_it}, we discuss the potential origin of the mixing in the stars of our sample and conclude in Sect. \ref{sec:conclusions} with future prospects and additional studies to be done. 

\section{Presentation of the sample}
\label{sec:sample}
We carried out a detailed modelling of 10 targets for which high quality asteroseismic data, spectroscopic constraint on lithium, surface rotation and Gaia parallaxes were available. We summarise these properties in Table \ref{tab:modelglobals}. The classical constraints used for the seismic modelling were $\rm{T_{eff}}$, L and $\left[ \rm{Fe/H} \right]$, except for the 16Cyg binary system were we favoured the interferometric radii values from \citet{White2013}, $\rm{R}_{A}=1.22 \pm 0.02 \rm{R}_{\odot}$ and $\rm{R}_{B}=1.12 \pm 0.02 \rm{R}_{\odot}$.

\begin{table*}[h]
\caption{Observational constraints for the stars under study.}
\label{tab:modelglobals}
  \centering
\begin{tabular}{l  c  c  c  c  c  c }
\hline \hline
Target& L $(\rm{L}_{\odot})$&$\rm{T_{eff}}$ (K)&$\left[ \rm{Fe/H} \right]$ ($\rm{dex}$)  & $\rm{A(Li)}$ ($\rm{dex}$) & Rotation Period $(\rm{P_{Rot}})$ $(\rm{d})$ & Ref. \\ \hline
KIC$5773345$&$5.33 \pm 0.19$&$6204 \pm 100$& $0.28 \pm 0.10$ & $2.3 \pm 0.1$& $11.29 \pm 0.62$ & 1 \\
KIC$5866724$&$2.68 \pm 0.10$&$6202 \pm 80$& $0.15 \pm 0.10$ & $2.95 \pm 0.21$ & $19.56_{-3.61}^{+2.24}$ & 2 \\ 
KIC$6521045$&$2.36 \pm 0.08$&$5795 \pm 80$& $ 0.05 \pm 0.10$ & $2.34 \pm 0.18$ & $22.35_{-2.18}^{4.17}$ & 2 \\ 
KIC$6679371$&$7.64 \pm 0.31$&$6379 \pm 100$& $0.00 \pm 0.10$ & $3.00 \pm 0.11 $ & $5.36 \pm 0.41 $ & 3 \\
KIC$7670943$&$2.78 \pm 0.10$&$6263 \pm 100$& $-0.08 \pm 0.10$ & $2.47 \pm 0.15$& $5.14 \pm 0.41$ & 2 \\
KIC$8292840$&$2.63 \pm 0.09$& $6357 \pm 100$& $-0.13 \pm 0.10$ & $2.66 \pm 0.13$ & $30.10_{-0.82}^{+0.54}$ & 2 \\
KIC$9410862$&$1.76 \pm 0.06$&$6137 \pm 150$& $-0.36 \pm 0.10$ & $2.6 \pm 0.16$& $22.63\pm 1.78$ & 3 \\
KIC$10079226$&$1.53 \pm 0.10$&$5943 \pm 100$& $0.16 \pm 0.10$ & $2.55 \pm 0.15$ & $15.51 \pm 2.35$ & 3 \\
16Cyg A &$1.544 \pm 0.081$&$5839 \pm 42$& $0.096 \pm 0.026$ & $1.30 \pm 0.08$& $23.80_{-1.8}^{+1.5}$ & 4 \\
16Cyg B &$1.228\pm 0.064$&$5809 \pm 39$& $0.052 \pm 0.021$ & $0.9$ & $23.20_{3.2}^{+11.5}$ & 4 \\
\hline
\end{tabular}
{\par\small\justify\textbf{Notes.} (1) Spectroscopic data from \citet{Pinsonneault2012} and seismic data from \citet{Lund2017}; (2) Spectroscopic data from \citet{Furlan2018} and seismic data from \citet{Davies2016}; (3) Spectroscopic data from \citet{Furlan2018} and seismic data from \citet{Lund2017};  (4) $T_{\mathrm{eff}}$ from \citet{White2013}, [Fe/H] from \citet{Ramirez2009}, $R$ from \citet{Metcalfe2012}, and seismic data from \citet{Davies2015}. \par}
\end{table*}

For 16Cyg A and B, we reused the set of constraints employed by \citet{Betrisey2023_AMS_surf}. The spectroscopic constraints for KIC5773345 were sourced from \citet{Pinsonneault2012}, while those for the remaining targets were obtained from \citet{Furlan2018}. The surface rotation periods obtained from photometric analyses were taken from \citet{Angus2018} and \citet{Santos2021}. For all targets, we estimated the absolute stellar luminosity using the following formula:
\begin{equation}
\log\left(\frac{L}{L_\odot}\right) = -0.4\left(m_\lambda + BC_\lambda -5\log d + 5 - A_\lambda -M_{\mathrm{bol},\odot}\right) \, ,
\label{eq_luminosity}
\end{equation}
where $m_\lambda$ is the magnitude and $\lambda$ is the wavelength band, in our case, the 2MASS $K_s$-band. The bolometric correction $BC_\lambda$ was computed according to \citet{Casagrande2014,Casagrande2018}. For consistency with their software, we adopted a solar bolometric correction of $M_{\mathrm{bol},\odot} = 4.75$. The distance $d$ in pc is based on \textit{Gaia} EDR3 \citep{Gaia2021}. In that regard, we tested two approaches: using the distance from \citet{Bailer-Jones2021} and inverting the parallax corrected according to \citet{Lindegren2021}. For the final luminosity, we adopted the former as both approaches led to consistent luminosities. The extinction $A_\lambda$ was inferred using the dust map from \citet{Green2018}. We used the seismic data from \citet{Davies2015} for 16Cyg A and B, from \citet{Davies2016} for KIC5866724, KIC6521045, KIC7670943, and KIC8292840, and from \citet{Lund2017} for the remaining targets. For 16Cyg A and B, we used in the modelling the interferometric radius provided by \citet{White2013}.

We adopted surface lithium abundances A(Li) from \citet{Gao2021Li}, based on a cross-match of the LAMOST survey with our \textit{Kepler} sample. For stars with multiple A(Li) measurements in that catalog (all except for KIC~5773345), we computed a weighted mean using the inverse squared uncertainty as weight. For 16Cyg A and B, we used the reported values in \citet{Morel2021}.

\section{Asteroseismic modelling using the FICO procedure}
\label{sec:asteroseimic_modelling_using_FICO_procedure}
The asteroseismic models were constructed using the Forward and Inversion COmbination (FICO) procedure \citep{Betrisey2022,Betrisey2023_AMS_surf,Betrisey2024_AMS_quality,Betrisey2024_phd}, which combines forward and inverse methods to achieve a detailed stellar characterization of solar-like stars, based on precursor works by \citet{Reese2012} and \citet{Buldgen2019f}. While mixing-length theory is a commonly used approximation for modelling convection \citep[e.g.][for textbooks]{Maeder2009,Kippenhahn2012}, including in our study, it has inherent limitations. This theory attempts to describe a complex, non-local 3D turbulent phenomenon using a simplified 1D local approximation, which inadequately represents the near-surface layers of stars \citep[e.g.][]{Nordlund&Stein1997,Stein&Nordlund1998,Tanner2013}. While this limitation is not critical for stellar physics overall, it becomes significant in asteroseismology because acoustic oscillations originate from these layers. Consequently, the inaccurate treatment of convection introduces biases in the theoretical frequency estimation \citep[e.g.][]{Kjeldsen2008,Ball&Gizon2014,Sonoi2015}.

The FICO procedure involves three main steps to construct a model that effectively mitigates these surface effects. From a general perspective, the FICO procedure employs frequency separation ratios to mitigate surface effects (third step). These ratios are specifically designed to minimise the impact of surface effects \citep{Roxburgh&Vorontsov2003}, albeit at the cost of information about the mean density. This information can be recovered quasi-model-independently through a mean density inversion (second step). We refer to \citet{Buldgen2022c} for a recent review about seismic inversions and briefly note that such techniques were successfully applied to various asteroseismic targets \citep[see e.g.][for a non-exhaustive list of references]{DiMauro2004_theo,Buldgen2016b,Buldgen2016c,Buldgen2017c,Buldgen2019b,Buldgen2019f,Kosovichev&Kitiashvili2020,
Salmon2021,Betrisey2022,Buldgen2022b,Betrisey2023_rot,Betrisey2023_AMS_surf,Betrisey2024_AMS_quality,
Betrisey2024_MA_Sun,Betrisey2025}. Since the mean density inversion relies on a linear formalism to deduce a correction on a so-called reference model \citep{Reese2012}, it must be performed on a model whose reference mean density is sufficiently close to the true mean density in the parameter space. Such a model is obtained by fitting the individual oscillation frequencies and applying a polynomial correction for the surface effects (first step). Although the stellar parameters of this reference model are slightly biased, the reference mean density is close enough to the true value for the inversion to provide a robust and reliable correction.

In practice, the following steps were undertaken. First, we employed the Asteroseismic Inference on a Massive Scale (AIMS) software \citep{Rendle2019} to fit the individual frequencies and spectroscopic constraints (in our case: effective temperature, metallicity, and luminosity). AIMS employs an MCMC-based algorithm, the Python \textsc{emcee} package \citep{Foreman-Mackey2013}, for interpolation between grid points of a precomputed grid of stellar models, and a Bayesian approach to derive posterior probability distributions of the optimised stellar parameters. The combination of a high-resolution grid with the interpolation scheme allows for a thorough exploration of the parameter space. For the minimisations of this section, we used an extended version of the Spelaion grid from \citet{Betrisey2023_AMS_surf}, whose physical ingredients are described in Appendix~\ref{app:spelaion_physics}. The Spelaion physics will be referred to as standard physics hereafter. Surface effects are modelled using the polynomial prescription of \citet{Ball&Gizon2014}. The following stellar variables are optimised in AIMS: mass, age, two mass fractions for the initial chemical composition ($X_0$ and $Z_0$), and the two free parameters for the surface effect prescription. For stars with masses exceeding $1.10M_\odot$, we also included the convective core overshooting parameter in the optimised variables. We employed uniform uninformative priors for all free variables, except for the stellar age, for which we assumed a uniform distribution between 0~and~13.8~Gyr. Observational constraints were assumed to be perturbed by Gaussian-distributed random noise for likelihood computations. The AIMS runs are executed in two phases. Initially, a burn-in phase identifies the relevant parameter space, followed by the solution run. For the first minimisation step, we used 800 walkers, 8000 burn-in steps, and 3000 solution steps, resulting in stellar parameters based on 2.4 million samples from the production run, following 6.4 million probability calculations from the burn-in phase. This extensive sampling was feasible due to the limited number of targets in our sample and was required to ensure robust convergence of the surface effect coefficients. Subsequently, a mean density inversion on this reference model is then carried out to get a robust constraint on the mean density, which is added to the set of constraints for a second MCMC fitting this time frequency separation ratios instead of individual frequencies. This second AIMS run is based on 2.4 million samples for both the production and solution runs.

Given the potential interpolation challenges when a target is near the end of the main sequence, a region characterised by higher degeneracies in the seismic information \citep[see e.g.][]{White2011}, we conducted an additional test to ensure the robustness of the final step of the FICO procedure. This test involved verifying that the interpolated frequencies from the AIMS solution can be reproduced by recomputing a stellar model using the AIMS optimal parameters and the same versions of the evolution and oscillation codes used for the stellar model grid. A successful interpolation would result in both frequency sets comparably reproducing the observational echelle diagram. Conversely, unsuccessful interpolation would lead to discrepancies in the echelle diagram, indicating that the AIMS interpolated solution is non-physical. Typically, this issue manifests as an overestimation of the stellar age, necessitating a local minimisation to achieve an unbiased stellar characterisation. Although rare, this scenario can occur in regions of the parameter space with higher degeneracy levels. We confirmed the robustness of the interpolation for most targets in our sample, with the exception of KIC6521045. For this target, we performed an additional local minimisation using a Levenberg-Marquardt algorithm \citep[e.g.][]{Frandsen2002,Teixeira2003,Miglio&Montalban2005}, employing the same free variables as in AIMS. The constraints included the $r_{01}$ frequency separation ratios, the spectroscopic constraints, the inverted mean density, and the lowest radial-order frequency. Since the Levenberg-Marquardt algorithm does not robustly propagate observational uncertainties, we utilised the uncertainties from AIMS. The uncertainty estimation in AIMS is reliable due to the thorough MCMC sampling of the parameter space around the solution, and we assumed that it was reasonable to consider the local minimisation solution sufficiently close to the AIMS solution for reusing this set of uncertainties. We used the asteroseismic models of 16Cyg A and B from \citet{Betrisey2023_AMS_surf}. These models were also constructed using the FICO procedure but a better set of constraints could be used for these targets. The asteroseismic modelling results are detailed in Table \ref{tab:modelFICO}.

\begin{table*}[h]
\caption{Stellar parameters determined from the FICO procedure using standard models.}
\label{tab:modelFICO}
  \centering
\begin{tabular}{l  c  c  c  c  c  c }
\hline \hline
Target& M $(\rm{M}_{\odot})$& R $(\rm{R}_{\odot})$& Age (Gyrs)& $X_{0}$  & $Z_{0}$ & $\alpha_{\rm{Ov}}$\\ \hline
KIC$5773345$&$1.582 \pm 0.026$&$2.040 \pm 0.013$& $2.14 \pm 0.11$ & $0.703$& $0.028$ & $0.19$\\
KIC$5866724$&$1.298 \pm 0.044$&$1.432 \pm 0.017$& $2.79 \pm 0.27$ & $0.697$ & $0.0257$ & $0.03$\\ 
KIC$6521045$&$1.122 \pm 0.080$&$1.511 \pm 0.008$& $7.38 \pm 0.12$ & $0.743$ & $0.0106$ & $0.18$\\ 
KIC$6679371$&$1.615 \pm 0.027$&$2.246 \pm 0.009$& $1.74 \pm 0.08$ & $0.686$ & $0.0247$ & $0.06$\\
KIC$7670943$&$1.235 \pm 0.041$&$1.420 \pm 0.016$& $3.52 \pm 0.33$ & $0.733$& $0.0154$ & $0.07$\\
KIC$8292840$&$1.140 \pm 0.030$&$1.332 \pm 0.012$& $2.75 \pm 0.25$ & $0.701$ & $0.0129$ & $0.00$\\
KIC$9410862$&$0.988 \pm 0.016$&$1.159 \pm 0.007$& $6.52 \pm 0.42$ & $0.738$& $0.0082$ & $0$\\
KIC$10079226$&$1.215 \pm 0.035$&$1.178 \pm 0.012$& $2.60 \pm 0.30$ & $0.741$ & $0.0209$ & $0.12$\\
16Cyg A &$1.057 \pm 0.008$&$1.216 \pm 0.005$& $7.01 \pm 0.10$ & $0.708$& $0.0198$ & $0$\\
16Cyg B &$1.010 \pm 0.024$&$1.105 \pm 0.009$& $7.06 \pm 0.11$ & $0.711$ & $0.0174$ & $0$\\
\hline
\end{tabular}
{\par\small\justify\textbf{Note.} Overshooting has been fixed to 0 for masses below $1.1\rm{M_{\odot}}$. \par}
\end{table*}

\section{Modelling including lithium depletion}
\label{sec:modelling_including_lithium_depletion}

The second part of our modelling intended at reproducing the observed lithium depletion in these targets while also remaining as consistent as possible with asteroseismic observations. To do this, we considered two distinct physical processes. The first one is turbulent diffusion, modelled as in \citet{Proffitt&Michaud1991} and using the following equation:
\begin{align}
D_{\rm{Turb}}=D_{T} \left(\frac{\rho_{\rm{BCZ}}}{\rho}\right)^{n} \label{eq:Proff}
\end{align}
with $\rho_{\rm{BCZ}}$ the density at the BCZ position, $\rho$ the local density, $D_{T}$ and $n$ being the free parameters in this formalism. This empirical approach defines ``turbulent'' diffusion as a sort of umbrella term that can be linked with various physical processes \citep[see][for a reference book on instabilities and transport processes]{Maeder2009}. In the following work, we test and compare the required efficiency to reproduce the observed depletion. We note that in the solar case, this has been calibrated to reproduce simultaneously the lithium and beryllium abundance \citep{Buldgen2025Li} and that attempts at directly linking the observed depletion to angular momentum transport processes have been made \citep{Eggenberger2022}, with the beryllium abundance playing a key discriminating role.

The second process is convective undershooting at the base of the envelope. The particularity of lithium is that it is not only burned during the main-sequence but also significantly during the pre main-sequence evolution. In the early evolutionary phases, the fusion point of lithium is in the convective envelope. As hydrogen burning starts, the convective envelope shrinks and excludes the fusion point of lithium, stopping the depletion. By adding convective undershooting at the base of a convective envelope, one may extend the duration over which the lithium fusion point remains in the convective envelope and the surface abundance is depleted. In our models, we follow a simple formalism of convective penetration for undershooting at the base of the envelope and will come back below on our motivations to do so. The equation describing the undershooting parameter is
\begin{align}
d_{\rm{Under}}=\alpha_{\rm{Under}}\times H_{P}, 
\end{align}
with $d_{\rm{Under}}$, the distance over which undershooting is acting, $\alpha_{\rm{Under}}$ is a free parameter similar to the mixing-length parameter and $H_{P}$ is the pressure scale height, defined as $H_{P}=-\frac{dr}{dln P}$. The mixing in this region is considered instantaneous and the temperature gradient is fixed to the adiabatic gradient. While other prescriptions exist and have been linked with hydrodynamical simulations \citep[e.g.][]{Viallet2015}, asteroseismic data has in some cases shown that strong discontinuities could also be observed at the base of convective envelopes \citep[e.g.][]{Lebreton2012}.

The ability of undershooting at the base of convective envelopes to be at the origin of the observed lithium depletion in solar-like stars has been recently put forward by \citet{Baraffe2017} and \citet{Thevenin2017} and has been mentioned in the solar case in various works \citep[e.g.][]{JCDOV,Zhang2019,Dumont2021,Dumont2021b}. 

Due to the lower precision of the lithium abundance determination compared to other observational constraints such as oscillation frequencies, the strategy we adopted was to carry out the modelling without including it explicitly in the cost function, but simply verifying a posteriori that our optimal asteroseismic model actually fell within the $1\sigma$ error bar. While slightly more tedious, this also makes the minimization procedure more stable given the potential degeneracies that may appear should extra-mixing be simply treated as a free parameter. This is a consequence of the fact that we model stars presenting only acoustic modes and that we apply a local minization technique based on a Levenberg-Marquardt algorithm. The potential weak sensitivity of the acoustic mode to extra-mixing (particularly in the form of turbulent diffusion) coupled to the computation of derivatives in the minimization strategy may lead to convergence issues. Nevertheless we are able to determine for each star a suitable model for the analysis of light element depletion using the aforementioned approach. We start by showing in Table \ref{tab:modeLM} the updated parameter values for these stars. As they have been estimated using a local minimization technique, we discard the uncertainties from the modelling as unreliable and note that the local solution we find would be revised with further revision in the physics. We also note that fitting the $r_{01}$ ratios in some cases led to more evolved solutions exhibiting negative frequency separation ratios typical of the subgiant branch evolution. These models were discarded as unreliable as they did not lead to a significantly improved fit but were more a symptom of convergence issues of the method. As we will see below, they may result from an attempt of the evolutionary models to fit the frequency separation ratios. 

\begin{table*}[h]
\caption{Stellar parameters determined from the local minimization including light element depletion.}
\label{tab:modeLM}
  \centering
\begin{tabular}{l  c  c  c  c  c  c  c  c  c  c }
\hline \hline
Target& M $(\rm{M}_{\odot})$& R $(\rm{R}_{\odot})$& Age (Gyrs)& $X_{0}$  & $Z_{0}$ & $\alpha_{\rm{MLT}}$& $\alpha_{\rm{Ov}}$ & $\alpha_{\rm{Under}}$& $D_{T}$ $(\rm{cm^{2}/s})$&$n$\\ \hline
KIC$5773345$&$1.510$&$2.032$& $2.41$ & $0.69$& $0.027$&$1.838$& $0.01$& $0.42$ & $10000$ & $2$\\
KIC$5866724$&$1.271$&$1.422$& $2.80$ & $0.707$ & $0.0215$& $1.701$& $0.06$& $0.15$ & $10000$ & $2$\\ 
KIC$6521045$&$1.110$&$1.506$& $ 7.18$ & $0.735$ & $0.0121$& $1.745$ & $0.0$ & $0.0$ & $10000$ & $2$\\ 
KIC$6679371$&$1.609$&$2.234$& $1.75$ & $0.682$ & $0.0244$& $1.781$ & $0.05$&  $0.35$ & $10000$ & $3$\\
KIC$7670943$&$1.236$&$1.429$& $3.68$ & $0.731$& $0.0152$& $1.973$&$0.05$ &$0.25$ & $10000$ & $3$\\
KIC$8292840$&$1.110$& $1.320$& $2.82$ & $0.677$ & $0.0122$& $1.687$& $0.0$&$0.30$ & $15000$ & $2$\\
KIC$9410862$&$0.993$&$1.160$& $6.61$ & $0.739$& $0.0083$& $1.884$& $0.0$&$0.0$ & $3000$& $3$\\
KIC$10079226$&$1.216$&$1.177$& $2.76$ & $0.742$ & $0.0218$& $1.850$ & $0.14$ & $0.0$ & $10000$ & $3$\\
16Cyg A & $1.059$ & $1.217$ & $7.11$ & $0.702$ & $0.0211$ & $1.843$ & $0.0$ & $0.0$ & $2600$ & $3$\\
16Cyg B & $1.027$ & $1.111$ & $7.05$ & $0.726$ & $0.0169$ & $1.842$& $0.0$& $0.0$ & $3600$ & $3$\\
\hline
\end{tabular}
\end{table*}

\begin{figure}
	\centering
		\includegraphics[width=8.5cm]{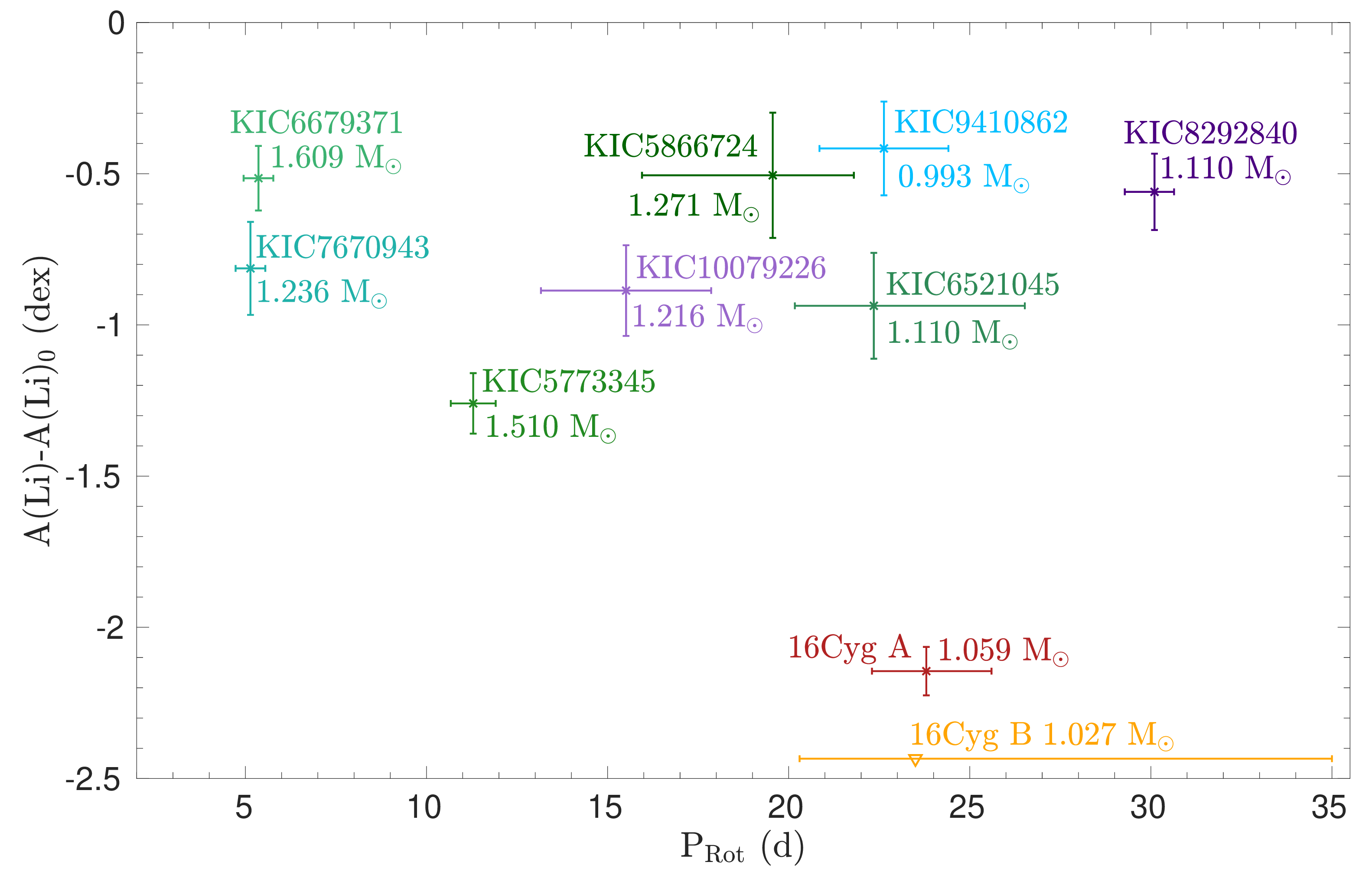}
	\caption{Surface lithium depletion as a function of surface rotation for all stars in our sample.}
		\label{Fig:LiDeplRot}
\end{figure}

Looking at Fig. \ref{Fig:LiDeplRot} and Table \ref{tab:modeLM}, we can examine a couple of trends. First, there does not seem to be a direct correlation of the lithium depletion (defined here as A(Li)-A(Li)$_{0}$, with A(Li)$_{0}$ the initial surface lithium abundance of the optimal model) with the rotational velocity, nor with mass. This is likely a result of the different spectral types in our sample that affect the rotational braking, as a clear correlation is found by \citet{Beck2017}. In this respect, our sample of stars that could be qualified of solar analogues is too small to draw any conclusions and a consistent modelling of a large sample should be made, which will likely possible thanks to the upcoming PLATO mission \citep{Rauer2024}. 

For stars such as 16Cyg A, 16Cyg B, KIC$9410862$ and KIC$6521045$, a classical, moderate turbulence at the BCZ can allow to reproduce the observed depletion. The efficiency of the turbulent mixing is about $50\%$ lower than the one found by \citet{Buldgen2025Li} to reproduce the solar lithium and beryllium values when no opacity increase is included. It is however in good agreement with the efficiency required in the solar case when an opacity increase at the BCZ is taken into account. This illustrates well the existing correlation between the position of the BCZ and lithium depletion in evolutionary models of solar-like stars and may also imply that some compensation is introduced when carrying asteroseismic modelling that limits the impact of the reference solar abundances. In the case of 16Cyg, this was already observed in \citet{Buldgen2016b,Farnir2020} and is further confirmed here. 

For all the other stars, despite the fact that the ad-hoc turbulence coefficient was set to quite high values  (namely $n=2$ or $3$ and $D_{T}=10000$ $\rm{cm^{2}/s}$, or in one case even $15000$ $\rm{cm^{2}/s}$), it was not possible to reproduce the depletion without undershooting at the base of the convective envelope. We mention that KIC$6521045$ could fall into this category and that its observed depletion could also easily be reproduced with adiabatic undershooting at the base of the convective envelope and a reduced efficiency of turbulent diffusion. 

\begin{figure*}
	\sidecaption
		\includegraphics[width=12cm]{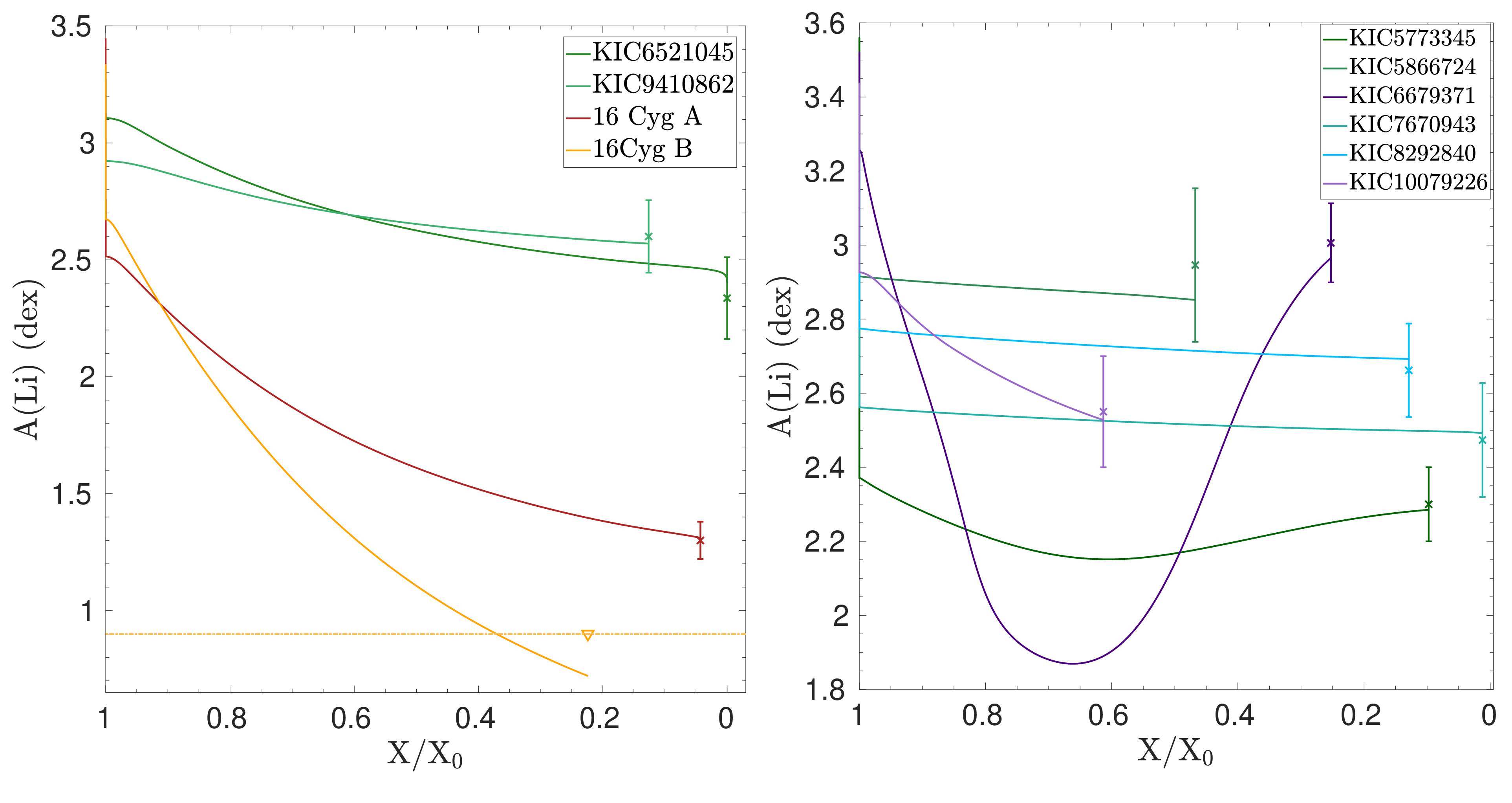}
	\caption{Left panel:.Evolution of the surface lithium abundance (A(Li)) as a function of central hydrogen abundance (normalised by the initial value) for the lower mass stars of our sample (G-type stars). The orange horizontal bar indicates the higher limit value for $16$Cyg B. Right panel: Same as the left panel for the higher mass stars of our sample (F-type stars) for which undershooting at the base of the convective envelope is necessary to reproduce the surface lithium abundance.}
		\label{Fig:LiDepleAll}
\end{figure*}

As can be seen from Fig. \ref{Fig:LiDepleAll}, this leads to a very different shape of the depletion, reminiscent of the plots in \citet{Thevenin2017}, whereas the depletion observed in models including only ad-hoc turbulence remains similar to the solar case, as illustrated in \citet{Buldgen2025Li}. A similar situation has been found in \citet{Moedas2025}, where they also found that turbulence in the radiative zone is not sufficient to explain the lithium depletion in their sample of \textit{Kepler} targets. We also note the much steeper depletion of 16Cyg B compared to 16Cyg A, which indicates either a very different rotational history or, as put forward by \citet{Deal2015}, the impact of a potential early accretion event of a planetary companion. In this case we find that the value of the calibrated turbulent coefficient is $35\%$ higher in 16Cyg B than in 16Cyg A.

\begin{figure*}
	\sidecaption
		\includegraphics[width=12cm]{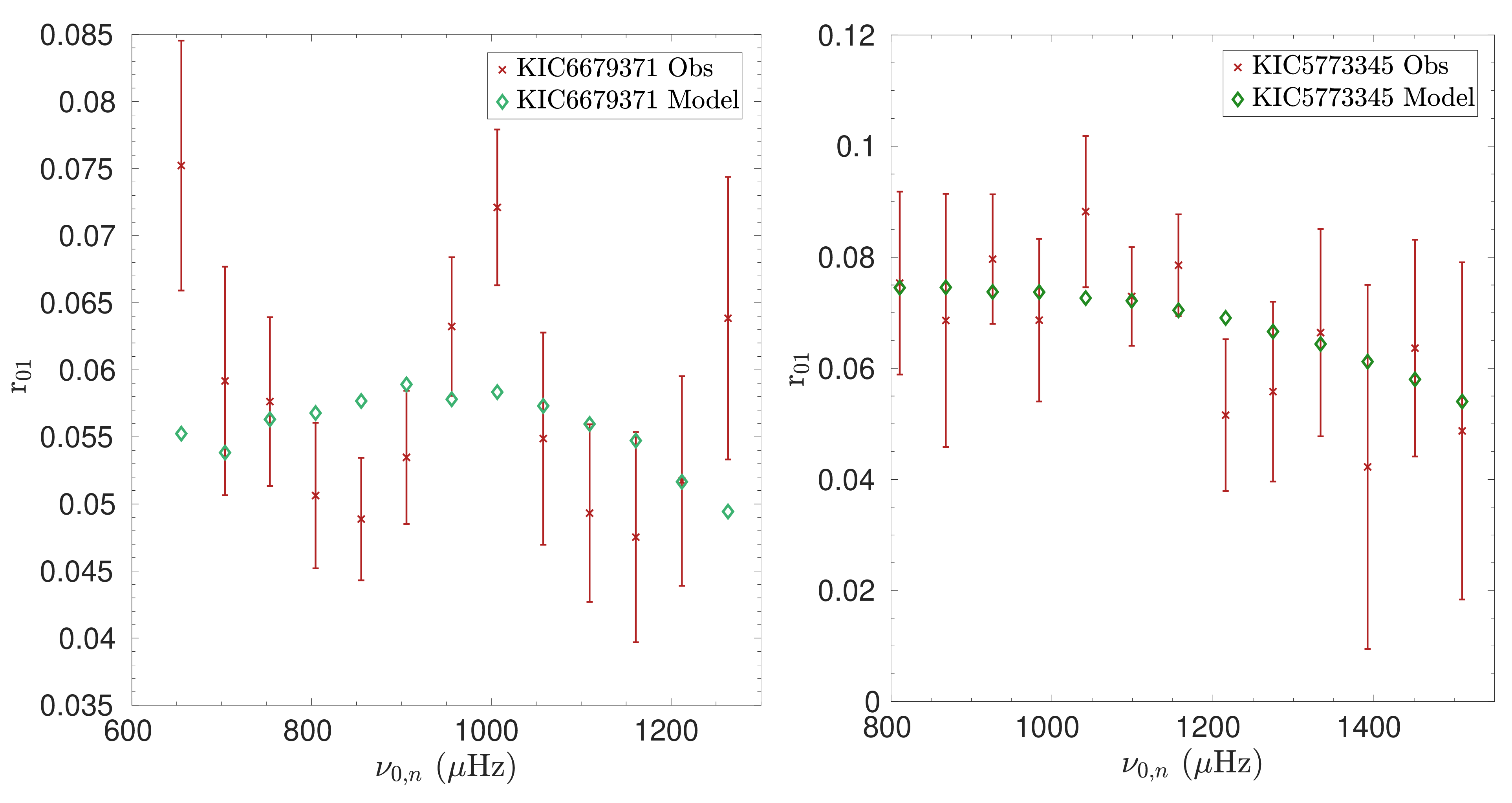}
	\caption{Frequency separation ratios for the optimal models found after the Levenberg-Marquardt minimization ($r_{01}$) of KIC$6679371$  (left) and KIC$5773345$ (right) as a function of observed frequency. The evolutionary model is shown in green while the observations are shown in red.}
		\label{Fig:r01FType}
\end{figure*}

\begin{figure*}
	\sidecaption
		\includegraphics[width=12cm]{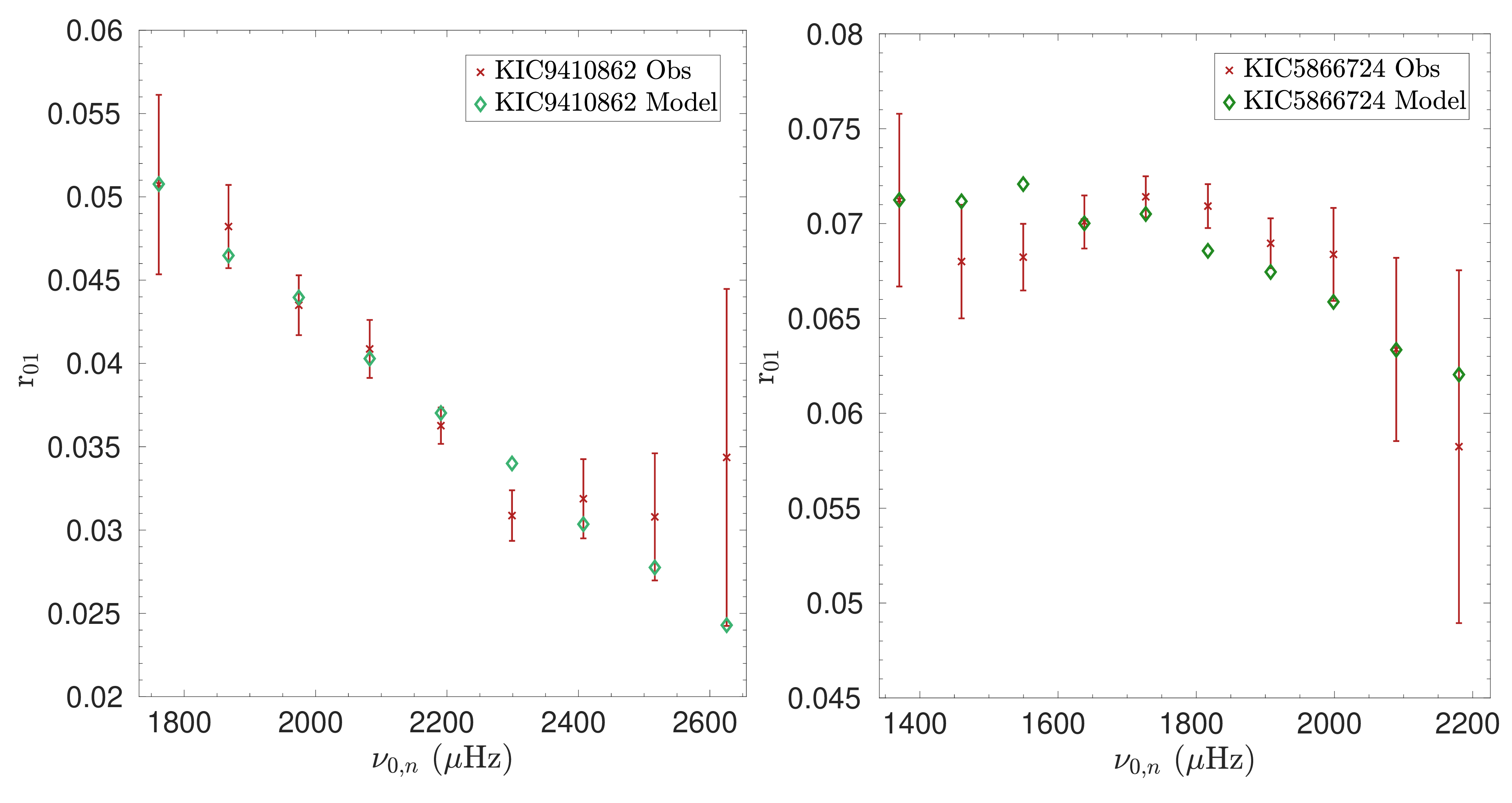}
	\caption{Frequency separation ratios for the optimal models found after the Levenberg-Marquardt minimization ($r_{01}$) of KIC $9410862$ (left) and KIC $5866724$ (right) as a function of observed frequency. The evolutionary model is shown in green while the observations are shown in red.}
		\label{Fig:r01GType}
\end{figure*}

Two stars, KIC$5773345$ and KIC$6679371$ show a peculiar behaviour in their lithium abundance due to their position at a later stage of hydrogen burning and a combined effect of turbulent diffusion and adiabatic undershooting at the base of the convective envelope. These stars have proven particularly difficult to fit and as shown in Fig. \ref{Fig:r01FType}, their $r_{01}$ ratios (as defined in \citealt{Roxburgh&Vorontsov2003}) are not at all reproduced. The situation is particularly dire for KIC$6679371$, despite the solutions being found to be similar to those of the FICO procedure and to those found in the literature \citep{Silva-Aguirre2017}. This is in strong contrast with lower mass G-type stars for which the $r_{01}$ can be matched by the evolutionary models, as shown in Fig. \ref{Fig:r01GType}. This indicates an apparent contradiction for both stars, as from their frequency separation ratios, one would expect a much deeper convective envelope, as confirmed by other studies  \citep{Lebreton2012, Deal2025}. However, the lithium depletion is quite moderate compared to the other stars and is reproduced with the calibrated value of undershooting at the base of the convective envelope used in the model. 

The higher abundance of lithium is a natural consequence of the evolution, as illustrated in Fig. \ref{Fig:LiDepleAll} for KIC$6679371$. We see that the deepening of the convective envelope actually leads to an increase of the abundance of lithium, as the convective layers reach regions that are enriched in lithium. Indeed, unlike the case of lower mass stars, a significant fraction of the radiative layers are too cold to undergo lithium burning. From this behaviour of the higher mass models close to the turnoff, a potential solution would be to use a slightly more evolved model. The main difficulty is then to keep the star on the main-sequence. In our tests, attempting to reproduce the $r_{01}$ value of KIC$6679371$ systematically induced our models to evolve on the subgiant branch and the ratios to become negative, even with high values of adiabatic undershooting at the base of the convective envelope. We come back to this potential solution in Sect. \ref{sec:origin_mixing_and_how_constrain_it}. This discrepancy can also be the trace of missing physical ingredients in the evolutionary models we have used, while other explanations may be linked with the actual non-linear response of the frequency separation ratios \citep{Deal2025}. In this case, our findings seem to indicate that even directly fitting the ratios without any analytical expression to reproduce the glitch does not allow to provide an acceptable agreement and the lithium abundance of these stars confirms that they should be close to the end of the main sequence. This work thus motivates further theoretical investigations to explain the observed differences and the reason why evolutionary models are unable to reproduce the data. 

\begin{figure}
	\centering
		\includegraphics[width=8cm]{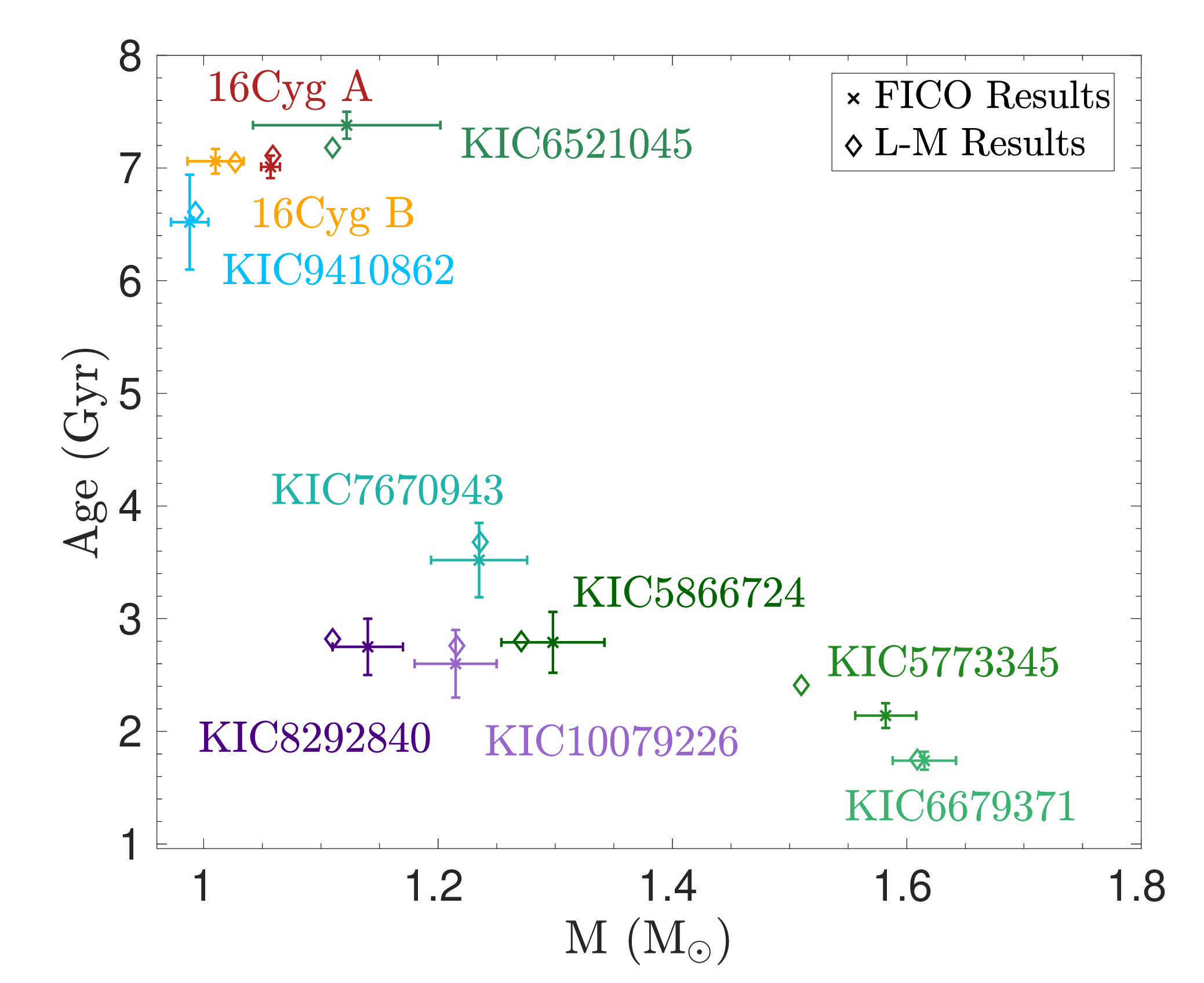}
	\caption{Variations in the determined asteroseismic masses and ages using the FICO procedure and a local Levenberg-Marquardt minimization that allowed to reproduce the observed lithium depletion.}
		\label{Fig:MAge}
\end{figure}

In our current modelling, the shifts in mass and age from the inclusion of the lithium depletion and variation of the transport of chemicals in the models remain minute, but further investigation is required as solutions aiming at reproducing the $r_{01}$ for KIC$6679371$ tend to go as low as $1.4$M$_{\odot}$, thus about a $15\%$ change. For solar-like stars, the changes remain minute and well within the requirements of the PLATO mission. This is illustrated in Fig. \ref{Fig:MAge} for all targets of the sample. A clear trend is observed with the more massive stars showing larger deviations. We stress that these values remain lower estimates of the possible shifts, given that the agreement with seismic indicators is not yet satisfactory in some cases and require further theoretical investigations. It is also likely that the effects of radiative accelerations or calibrated values of convective core overshooting from numerical simulations will impact the higher mass solutions, particularly KIC$6679371$ and KIC$5773345$.

Comparing the variations in Fig. \ref{Fig:MAge} however only tells one part of the story. Looking at the $\alpha_{\rm{MLT}}$ or $\alpha_{\rm{Ov}}$ parameters in Tables \ref{tab:modelFICO} and \ref{tab:modeLM}, we find more significant discrepancies between the two solutions. It is however difficult quantitavely conclude on their validity, while departures from the solar calibrated value of $\alpha_{\rm{MLT}}$ used for the grid in the FICO procedure are expected, we cannot safely exclude compensations. For example, it is likely that the reduction in the efficiency of settling resulting to the added turbulence in the radiative zone to reproduce lithium induced a variation of $\alpha_{\rm{MLT}}$ and $\alpha_{\rm{Ov}}$ to restore the agreement with seismic and classical constraints. Further tests using entropy-calibrated as in \citet{Manchon2024} might help although will unlikely explain the discrepancies discussed above for KIC$6679371$ and KIC$5773345$.

\section{Attempting to reconcile the frequency separation ratios and the lithium abundance}
\label{sec:origin_mixing_and_how_constrain_it}

\begin{figure}
	\centering
		\includegraphics[width=8cm]{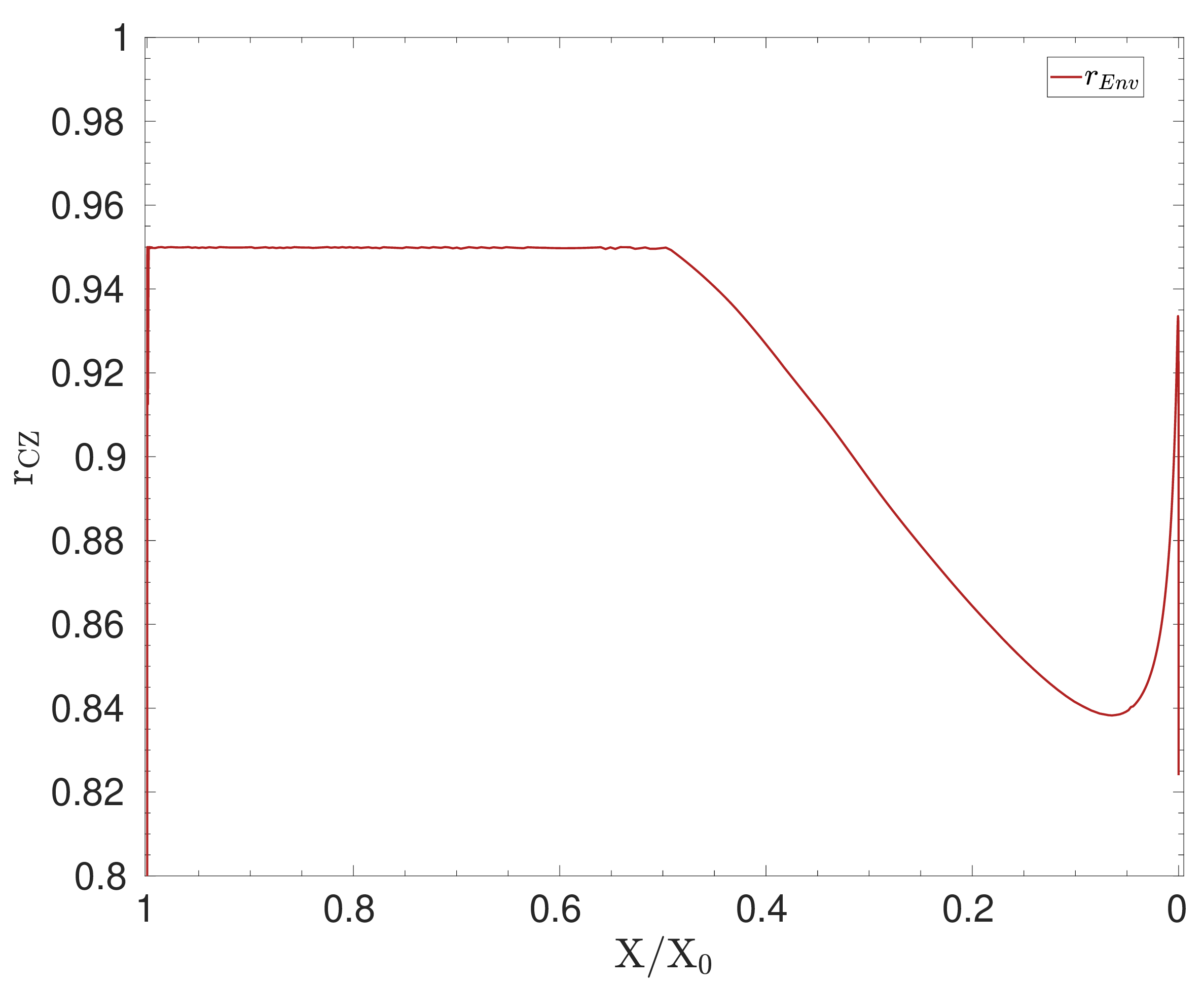}
	\caption{Evolution of the position of the base of the envelope, $r_{Env}$ as a function of normalized central hydrogen abundance for the model of KIC$6679371$.}
		\label{Fig:RCZ}
\end{figure}

As mentioned above, the situation observed in the case of KIC$6679371$ and KIC$5773345$ is quite peculiar, as the BCZ position will naturally extend deeper in the model while the lithium abundance will increase and finally form a plateau. This is linked to the convective envelope reaching deeper regions in the star, yet unaffected by lithium combustion. This is illustrated in Fig. \ref{Fig:RCZ}, where we show that towards the second half of the MS, the envelope deepens as the convective core shrinks. This is in line with the reincrease of the lithium that we see in the right panel of Fig. \ref{Fig:LiDepleAll}. We note however that there is an effect of turbulent diffusion, as both the decrease and increase of the surface lithium abundance are not fully linked with the BCZ position. Indeed, significant burning still occur during the MS under the effect of turbulent diffusion, that is enhanced by the extension of the convective envelope using undershooting at its lower boundary.

\begin{figure*}
	\centering
		\includegraphics[width=15cm]{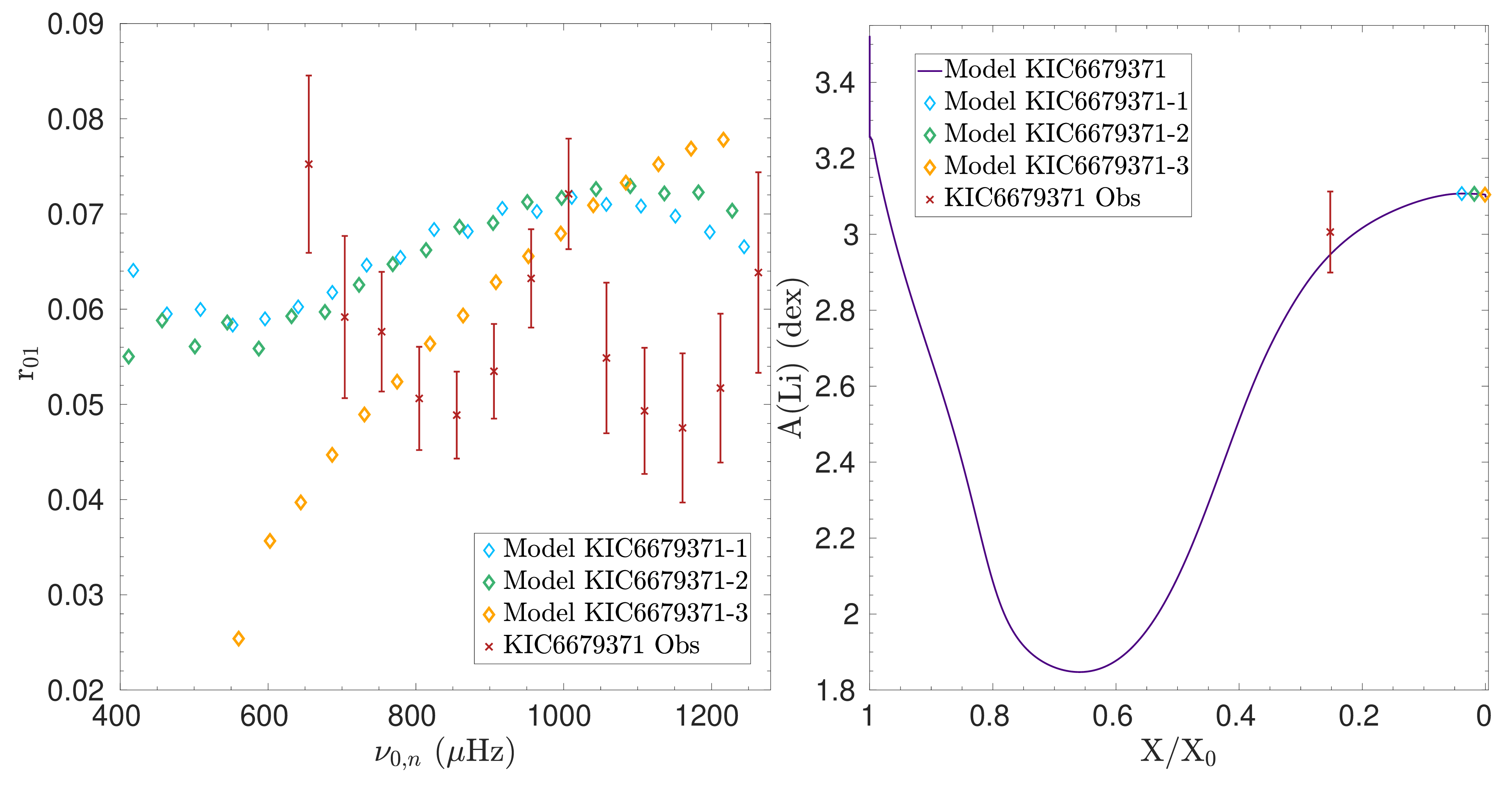}
	\caption{Left panel: Frequency separation ratios ($r_{01}$) of KIC$6679371$ as a function of observed frequency. Right panel: Surface lithium abundance as a function of normalised central hydrogen abundance. The observed value, associated with the $X/X_{0}$ from the optimal model, is shown in red. The green, orange and blue diamonds indicate the positions of Model 1, 2 and 3.  Three evolutionary models at the end of the main-sequence (noted Model 1, 2 and 3) are shown. }
		\label{Fig:RatiosEvol}
\end{figure*}

We show in Fig. \ref{Fig:RatiosEvol} the evolution of the surface lithium abundance for a model evolved beyond the ``optimal age'' found for KIC$6679371$ as well as a few examples of frequency separations ratios $r_{01}$ for the observed modes. Given that it forms a plateau towards the end of the main-sequence, lithium is essentially uninformative on the actual age of the star as soon as the convective envelope expands. Regarding the frequency separation ratios, illustrated in the left panel of Fig. \ref{Fig:RatiosEvol}, we can see that letting the model evolve further is not the solution, at least not at fixed mass. None of the models exhibit the high-amplitude glitches observed in the $r_{01}$ ratios of KIC$6679371$ despite the BCZ being $5\%$ deeper than in the optimal model, which appears largely insufficient. Moreover, the slope of the $r_{01}$ ratios as a function of frequency remains sensitive to the core properties, we can observe a rapid and drastic increase, in contradiction with the behaviour in the observations and in main-sequence models that exhibit either a rather flat frequency dependency or a negative slope of the $r_{01}$ separation ratios. This is mostly seen in stars close to the subgiant branch and it is expected that the $r_{01}$ will take negative values \citep{White2011, Deheuvels2016}. This was actually seen for the third model that exhibits a steep slope in the low frequency regime due to the core contraction in this model.

This implies that on top of the BCZ having to be significantly deeper, the core properties, to which the slope of the $r_{01}$ ratios remains sensitive, must be those of a main-sequence star. This further complicates the analysis and requires more detailed investigations likely with a different optimization technique than the Levenberg-Marquardt approach used here.

\section{Conclusions}
\label{sec:conclusions}

In this paper, we have carried out a detailed modelling procedure of 10 targets of the \textit{Kepler} Legacy sample for which a spectroscopic measurement of the surface lithium abundance was available. We have presented in Sect. \ref{sec:sample} the sample of stars we studied and the determination of the luminosity using spectroscopic constraints and Gaia parallaxes. In Sect. \ref{sec:asteroseimic_modelling_using_FICO_procedure}, we have presented the results obtained from the modelling using the FICO procedure \citep{Betrisey2023_AMS_surf}. These values constitute updated references parameters for these targets using the most recent modelling strategies and updated physical ingredients regarding previous studies in the literature. However, as they did not include the lithium depletion, we carried out a dedicated modelling of the targets using asteroseismic constraints and aiming at adding either turbulence or undershooting at the base of the convective envelope to reproduce the observed lithium depletion. The results of this modelling were presented in Sect. \ref{sec:modelling_including_lithium_depletion} and show in general relatively minor modifications from the ones obtained from the FICO procedure. From this analysis, we found that for the G-type, lower mass stars in our sample, the lithium depletion could easily be reproduced by adding a turbulent diffusion at the base of the convective envelope, similarly to what is done in the solar case. Meanwhile, for the more massive stars of our sample, even pushing to higher values of this ad-hoc correction was insufficient to reduce the surface lithium abundance. We found that this is a direct consequence of the very shallow convective zone these stars exhibit, meaning that extending the convective envelope was required to reproduce the observed lithium abundance. 

By adding undershooting at the base of the convective envelope, we are able to deplete lithium in the early main-sequence and reproduce the observed values. While this provides a good fit of the $r_{01}$ frequency separation ratios, we find that in some specific cases, namely KIC$6679371$ and KIC$5773345$, the lithium abundance is too high with respect to the required undershooting at the base of the convective envelope to reproduce the frequency separation ratios. While this is not in disagreement with evolutionary model predictions, as the lithium abundance in this mass range is expected to reincrease at the end of the main-sequence, the observed signature in the ratios cannot be reproduced by simply pushing the models to a slightly higher age, as shown in Sect. \ref{sec:origin_mixing_and_how_constrain_it}. Our results confirm the investigations of \citet{Deal2023} and \citet{Deal2025} and may imply that either the evolutionary solution is not reachable in our models because of missing physical ingredients or the behaviour of the modes is not properly described with our oscillation codes. The modelling of convection by the mixing length theory used in the evolutionary models might not be suitable in such thin convective envelopes. It is therefore possible that the observed behaviour here for pressure modes is a result of a stronger coupling between oscillations and convection. Issues with the treatment of time-dependent convection and turbulence models have also been cited as explanations for the difficulties in reproducing the $\gamma$ Doradus instability strip \citep{Dupret2005} which lie in a similar mass range as KIC$6679371$ and KIC$5773345$. This also implies that lithium might not be a good calibrator of turbulent mixing in F-type stars as its depletion could be due to, or even requires, undershooting at the base of the convective envelope. Fortunately, signatures on the helium abundance in the convective envelope may provide a calibration of turbulent transport \citep{Verma2019}. Further theoretical investigations on the behaviour of the frequency separation ratios are required to conclude in one direction or the other. This however puts some suspicions on the inferred masses and ages in this mass range and especially the values of KIC$6679371$ and KIC$5773345$ reported here. 

Future investigations are required to optimally exploit the seismic data provided by the PLATO mission \citep{Rauer2024} and to define robust modelling strategies for these targets that may be a significant fraction of the main PLATO sample \citep{Goupil2024}. At first, a better understanding of the observed asteroseismic signatures, in line with the works of \citet{Deal2025} will be crucial to interpret these observations. Finally, our findings also imply that testing fundamental physics ingredients might require more advanced numerical schemes than Levenberg-Marquardt algorithms, even though a first round of MCMC modelling has been carried out. This is particularly true for stars close to the turn-off that may display strong non-linear behaviours.


\begin{acknowledgements}
We thank the referee for their careful reading of the manuscript and insightful comments. G.B. acknowledges fundings from the Fonds National de la Recherche Scientifique (FNRS) as a postdoctoral researcher. J.B. acknowledges funding from the SNF Postdoc.Mobility grant no. P500PT{\_}222217 (Impact of magnetic activity on the characterization of FGKM main-sequence host-stars). We are grateful to Saniya Khan for her help with the determination of the luminosity and the use of the extinction and bolometric correction softwares. We thank Morgan Deal for the fruitful discussions on the issues linked with the ratios of F-type stars and particularly of KIC$6679371$.
\end{acknowledgements}


\bibliography{bibliography.bib}

\appendix

\section{The Spelaion physics}
\label{app:spelaion_physics}
The Spelaion grid comprises several subgrids, each designed to model specific types of stellar physics \citep[see][]{Betrisey2023_AMS_surf}. In Sect.~\ref{sec:asteroseimic_modelling_using_FICO_procedure}, we utilised two subgrids, labelled as `Standard MS' and `Overshooting MS' in \citet{Betrisey2023_AMS_surf}. The Standard MS subgrid has a convective core overshooting parameter fixed at zero, while the Overshooting MS subgrid includes the core overshooting parameter as a free variable. We note that these two subgrids have been extended to cover slightly larger masses and lower metallicities since the publication of the \citet{Betrisey2023_AMS_surf} paper, which is why we refer to the extended Spelaion grid in the main text.

The evolutionary sequences of the subgrids were computed using the Liège evolution code \citep[CLES;][]{Scuflaire2008b}. For each model in the sequence, the adiabatic Liège oscillation code \citep[LOSC;][]{Scuflaire2008a} was used to compute the acoustic oscillation frequencies. The stellar models incorporate AGSS09 abundances (Asplund et al. 2009) and OPAL opacities \citep{Iglesias&Rogers1996}, supplemented by \citet{Ferguson2005} opacities at low temperatures and electron conductivity by \citet{Potekhin1999}. The equation of state is described by the FreeEOS equation of state \citep{Irwin2012}, and it is worth noting that the OPAL opacities are based on the OPAL2005 equation of state \citep{Rogers&Nayfonov2002}. We employed the \citet{Thoul1994} formalism for microscopic diffusion, with the screening coefficients of \citet{Paquette1986}. The nuclear reaction rates were sourced from \citet{Adelberger2011}. Convection was modelled using the mixing-length theory from \citet{Cox&Giuli1968}, with a mixing-length parameter $\alpha_{\mathrm{MLT}}$ fixed at a solar-calibrated value of 2.05. We used the $T(\tau)$ relation from model~C of \citet{Vernazza1981} for atmospheric modelling.

\end{document}